\documentstyle[aps,multicol,epsf]{revtex}
\begin{document}
\draft
\title{Diffusion of small light particles in a solvent of large massive molecules}

\vspace{1cm}
\author{Rajesh K. Murarka, Sarika Bhattacharyya\footnote[2]{Present Address: Arthur Amos Noyes Laboratory of Chemical Physics, California Institute of Technology, Pasadena, California 91125}and Biman Bagchi\footnote[1]{For correspondence: bbagchi@sscu.iisc.ernet.in}}
\address{Solid State and Structural Chemistry Unit,\\
Indian Institute of Science,\\
Bangalore, India 560 012.}
\maketitle

\begin{abstract}
  
We study diffusion of small light particles in a solvent which consists
of large heavy particles. The intermolecular interactions are chosen to 
approximately mimic a water-sucrose (or water-polysaccharide) mixture. 
Both computer simulation and mode coupling theoretical (MCT) calculations 
have been performed for a solvent-to-solute size ratio five and for a 
large variation of the mass ratio, keeping the mass of the solute fixed. 
Even in the limit of large mass ratio the solute motion is found to remain 
surprisingly coupled to the solvent dynamics. Interestingly, at intermediate values
of the mass 
ratio, the self-intermediate scattering function of the solute, $F_{s}(k,t)$ (where
$k$ is the wavenumber and $t$ the time), develops a 
stretching at long time which could be fitted to a stretched exponential 
function with a k-dependent exponent, $\beta$. 
For very large mass ratio, we find the existence 
of two stretched exponentials separated by a power law type plateau. 
The analysis of the trajectory shows the coexistence of both hopping and 
continuous motions for both the solute and the solvent particles. 
It is found that for mass ratio five, the MCT calculations of the self-diffusion 
underestimates the simulated value by about $20 \%$, which appears to be 
reasonable because the conventional form of MCT does not include the hopping 
mode. However, for larger mass ratio, MCT appears to breakdown more severely. 
The breakdown of the MCT for large mass ratio can be connected to a similar breakdown 
near the glass transition.
\end{abstract}

\begin{multicols}{2} 
\section{Introduction}

  The issue of diffusion of small light particles in a solvent composed of 
larger and  heavier particles is unconventional because the role of the 
solvent in the solute diffusion is different from the case where the sizes are 
comparable. There are two limits that can be identified for such systems. 
One limit is the well studied Lorentz gas system, which consists 
of a single point particle moving in a triangular array of immobile disk 
scatters. Here the motion of the point particle can be modelled by random walk 
between traps.$^{1-3}$ The other limit is where the size of the 
solute particle is still smaller than that of the solvent molecules but it has a 
finite size (that is, not a point) and while the solvent is slow (compared to the 
solute particles) but not completely immobile. In the latter case, the translational 
diffusion of the solute is often attempted to describe by the well known 
hydrodynamic Stokes-Einstein (SE) relation given by,$^{4,5}$

\begin{equation}
D=\frac{k_{B}T}{C \eta R}
\end{equation}
\noindent
where $k_{B}$ is the Boltzmann constant, $T$ is the absolute temperature, 
C is a numerical constant determined by the hydrodynamic boundary condition, 
$\eta$ is the shear viscosity of the solvent and $R$ is the radius of the 
diffusing particle. Validity of Eq. 1 for small solutes is, of course, 
questionable.$^{4-6}$

 There have been many experimental,$^{6,7}$ 
computer simulation$^{8-10}$ and theoretical$^{11,12}$ 
studies of diffusion of small solute particles in a solvent composed of larger 
particles. All these studies show that the SE relation significantly 
underestimates the diffusion coefficient. To explain the enhanced diffusion, sometimes 
an empirical modification of the SE relation is used.$^{6,7}$ It is 
considered that $D \propto \eta^{-\alpha}$, where $\alpha \simeq 2/3$. This fractional 
viscosity dependence is referred to as the microviscosity effect which implies 
that the viscosity around the small solute is rather different from that of the 
bulk viscosity. The enhanced diffusion value has also been explained in terms of 
effective hydrodynamic radius.$^{4,13}$

 The earlier mode coupling theoretical (MCT) studies$^{11,12}$ of diffusion
of smaller solutes in a solvent composed of larger size molecules attributed 
the enhanced diffusion to the decoupling of the solute motion from the structural 
relaxation of the solvent. The MCT studies suggest that this decoupling of
the solute motion from the structural relaxation of the solvent can lead to the 
fractional viscosity dependence often observed in supercooled liquids. 

 However, there have been no systematic studies of the effects of the
variation of size and mass of the solute-solvent system. In this article 
we have explored the diffusional mechanism of the isolated small 
particles (solute) in a liquid composed of larger particles (solvent), both 
analytically and numerically. The study is performed by keeping the solvent-to-solute 
size ratio ($S_{R}$) fixed at five, but varying the mass of the solvent over a large 
range, by keeping the mass of the solute fixed. That is, the mass ratio $M_{R}$ 
(solvent mass/solute mass) is progressively raised to higher values. This system is 
expected to mimic some aspects of the water-sucrose or water-polysaccharide mixtures.$^{14}$
 
The trajectories of the solute and the solvent show coexistence of both hopping 
and continuous motions. As the solvent mass is increased, the self-intermediate scattering 
function of the solute develops an interesting stretching in the long time. For larger 
mass ratio we see the existence of two stretched exponentials separated by a power law 
type plateau.
 
 The mode coupling theory calculation of the self-diffusion coefficient of 
solute particles performed in the limit of small mass ratio (of 5) is found to be 
in qualitative agreement with the simulated diffusion -- MCT underestimates 
the diffusion by about $20 \%$. Thus, although the MCT underestimates the 
diffusion, the agreement is satisfactory in light of the contribution from the
hopping mode to diffusion which MCT does not explicitly take into account. However, the deviation 
from the simulated value increases with increase in mass ratio (which is equivalent 
to the increase of the mass of the solvent particles). In the limit of large 
mass ratio, MCT breaks down. The binary contribution to the total friction is found to 
{\it decrease} as one increases the mass of the solvent. In addition, due to the development 
of stretching in the self-intermediate scattering function of the solute and the inherent
slow solvent dynamics, there remains a strong coupling of the solute 
motion to the solvent density mode. This enhanced coupling at larger mass ratio
came as a surprise to us.

 In the limit of very large mass ratio, the motion of the light solute particle 
resembles to that of its motion in an almost frozen disorder system like near 
the glass transition temperature. Thus, the breakdown of MCT in the limit of large 
mass ratio could be connected to its failure near the glass transition temperature. 
Of course, one should note that in the limit of mass of the solvent goes to 
infinity, the basic assumption of conventional MCT breaks down.

It is widely believed that in a deeply supercooled liquid close to its glass
transition temperature ($T_g$), the hopping mode is the dominant mode in
the system which controls the mass transport and the stress relaxation. Recently,
a computer simulation study of a deeply supercooled binary mixture$^{15}$ has shown 
the evidence of an intimate connection between the anisotropy in local stress and the 
particle hopping. It was shown that the local anisotropy in the stress is responsible for 
the particle hopping in a particular direction. Furthermore, it was suggested that the local 
frustration present in the system (which is more in a binary mixture with components 
of different sizes) could cause the local anisotropy in the stress which in turn acts as a
driving force for hopping. However, in the present study, the density (or the pressure) of the 
system is not as high as that of a deeply supercooled system. The relaxation of the
stress is found to occur much faster and it relaxes almost completely within our simulation
time window even for the largest mass ratio. Consequently, the microscopic origin of particle 
hopping here could be different than that for a deeply supercooled liquid.

 The layout of the rest of the paper is as follows. Section II deals with 
the system and simulation details. The simulation results and discussions are 
given in the next Sec. III and the mode coupling theoretical analysis 
is presented in Sec. IV. In Sec. V, we discuss the possible effect of
heterogeneity probed by the solute on the self-dynamic intermediate scattering
function of the solute. Finally, in Sec. VI we present the 
conclusions of the study.

\section {System and Simulation Details}

We performed a series of equilibrium isothermal-isobaric (N P T) ensemble molecular 
dynamics (MD) simulation of binary mixtures in three dimensions for an infinitesimal 
small value of the mole fraction of one of the species. The binary system studied 
here contains a total of $N = 500$ particles consisting of two species of 
particles, with $N_1 = 490$ and $N_2 = 10$ number of particles. Hereafter, we refer 
the indices 1 and 2, respectively, for the solvent and solute particles. Thus, the 
mixture under study contains $2 \%$ of solute particles. The interaction between 
any two particles is modeled by means of shifted force Lennard-Jones (LJ) pair 
potential,$^{16}$ where the standard LJ is given by
\begin {equation}
u_{ij}^{LJ} = 4 \epsilon _{ij} \left[{\left(\sigma_{ij} \over r_{ij}\right)}^{12} - {\left(\sigma_{ij} \over r_{ij} \right)}^6 \right]
\end {equation}
\noindent where $i$ and $j$ denote two different particles (1 and 2). 
In our model system, the potential parameters are as 
follows: $\epsilon_{11} = 1.0$, $\sigma_{11} = 1.0$, $\epsilon_{22} = 1.0$, 
$\sigma_{22} = 0.2$, $\epsilon_{12} = 2.0$ (enhanced attraction), and $\sigma_{12} = 0.6$. 
The mass of the solute particles is chosen to be $m_2 = 0.2$ where the solvent (species 1) 
mass $m_1$ is increasingly varied and four different values are chosen 1, 5, 10 and 50. 
Thus, in this study we examined four different solvent-to-solute mass 
ratio, $M_{R}$ = $m_1/m_2$ = 5, 25, 50 and 250 for a fixed solvent-to-solute size 
ratio, $S_{R}$ = $\sigma_{11}/\sigma_{22}$ = 5. Note that in the model system being 
studied the solute-solvent interaction ($\epsilon_{12}$) is much stronger than 
both of their respective pure counterparts. In order to lower the computational 
burden the potential has been truncated with a cutoff radius of 2.5$\sigma_{11}$. 
All the quantities in this study are given in reduced units, that is, length in 
units of $\sigma_{11}$, temperature $T$ in units of $\epsilon_{11}/k_B$, pressure $P$ 
in units of $\epsilon_{11}/\sigma_{11}^{3}$, and the mass in the unit of $m$, which can
be assumed as argon (Ar) mass unit. The corresponding microscopic time scale is 
$\tau = \sqrt{m\sigma_{11}^2/\epsilon_{11}}$.

 All simulations in the NPT ensemble were performed using the 
Nose-Hoover-Andersen method,$^{17}$ where the external reduced temperature 
is held fixed at $T^* = 0.8$. The external reduced pressure has been kept fixed 
at $P^* = 6.0$. The reduced average density ${\bar \rho}^*$ of the 
system corresponding to this thermodynamic state point is 0.989 for all the mass 
ratios being studied. Throughout the course of the simulations, the 
barostat and the system's degrees of freedom are coupled to an independent 
Nose-Hoover chain$^{18}$ (NHC) of thermostats, each of length 5. 
The extended system equations of motion are integrated using the reversible 
integrator method$^{19}$ with a small time step of $0.0002$. The higher order 
multiple time step method$^{20}$ has been employed in the NHC evolution operator 
which lead to stable energy conservation for non-Hamiltonian dynamical systems.$^{21}$ 
The extended system time scale parameter used in the calculations was taken to be $0.9274$ 
for both the barostat and thermostats.   
   
 The systems were equilibrated for $10^6$ time steps and simulations were carried 
out for another $2.0\times 10^6$ production steps following equilibration, during 
which the quantities of interest are calculated. The dynamic quantities are averaged 
over three such independent runs for better improvement of the statistics. We have also 
calculated the partial radial distribution functions ($g_{11} (r)$ and $g_{12} (r)$) to 
make sure that there is no crystallization.

\section{Simulation Results and Discussion} 

In figure 1 we show typical {\it solute} trajectories for four different 
solvent-to-solute mass ratio, $M_{R}$ ($ = m_1/m_2$, $m_1$ is the mass of the 
solvent particles). The trajectories reveal interesting dependence on $M_{R}$. 
At the value of $M_{R}$ equal to five, the solute trajectory is mostly continuous 
with occasional hops. As the mass ratio $M_{R}$ is increased, the solute motion 
gets more trapped and its motion tends to become discontinuous where displacements 
occur mostly by hopping. This is because with increase in the solvent mass 
the time scale of motion of the solvent particles become increasingly slower. 
Thus, the solute gets caged by the solvent particles and keeps rattling between 
a cage till one solvent particle moves considerably to disperse the solute 
trajectory (see trajectory for $M_{R}$ = 250). Thus, there is a remarkable change 
in solute's motion in going from $M_{R}$ = 5 (figure 1a) to $M_{R}$ = 250 (figure 1d). 
  
In figure 2 we plot the {\it solvent} trajectories for different solvent-to-solute mass 
ratio, $M_{R}$. We find that for all values of $M_{R}$, there is a coexistence of 
hopping and continuous motion of the solvent molecules. At higher solvent mass, as 
expected, the magnitude of displacement becomes less and hopping becomes less 
frequent, but, surprisingly, the jump motion still persists.

Figure 3 displays the decay behavior of the self-intermediate scattering function ($F_{s}(k,t)$)
of the solute for different mass ratio $M_{R}$, at reduced wavenumber 
$k^* = k\sigma_{11} \sim 2 \pi$. The plot shows that $F_{s}(k,t)$ begins to 
stretch more for higher solvent mass. This stretching of $F_{s}(k,t)$ is kind 
of novel and we have examined it in detail.

After the initial Gaussian decay, $F_{s}(k,t)$, for smaller values of 
$M_{R}$, can be fitted to a single stretched exponential where the 
exponent $\beta \simeq 0.6$. However, for higher mass of the solvent, $F_{s}(k,t)$ 
can be fitted only to a sum of {\it two} stretched exponentials.
 
The behavior of $F_{s}(k,t)$ of the {\it solvent} for all the solvent masses studied is 
shown in figure 4. The plot shows that (as expected) the time scale of relaxation 
of $F_{s}(k,t)$ becomes longer as the solvent mass is increased. 
However, the self-intermediate scattering function of the solvent does not display any 
stretching at long times, even for the largest mass ratio considered. The decay 
can be fitted by sum of a short-time Gaussian and a long-time exponential function. 

The reason that $F_{s}(k,t)$ of the solute shows such stretching but that
of  the solvent does not, can be explained as follows. Due to the small size 
and the lighter weight of the solute, the time scale of motion of the solute particles
is much shorter compared to that of the solvent particles. Consequently, the solute 
motion probes more heterogeneity during the time scale of decay of $F_{s}(k,t)$ of
the solute. This heterogeneity probed by the solute increases as the solvent mass 
is increased. Since the solvent motion is much slower, it probes enough configurations 
during the time scale of decay of its $F_{s}(k,t)$.

In order to quantify the degree of heterogeneity probed by the solute, 
we have plotted the non-Gaussian parameter $\alpha_{2}(t)$,$^{22}$ for the solute, in 
figure 5. Clearly, the heterogeneity probed by the solute (quantified by the peak height 
of $\alpha_{2}(t)$) increases as the solvent mass is increased. On the 
other hand, $\alpha_{2}(t)$ of the solvent shows no such increase in the peak height 
of $\alpha_{2}(t)$ which remains small and unaltered, although the position of the peak 
shifts to longer time as the mass of the solvent is increased. We have discussed this 
analysis in more detail in section V.

The role of the local heterogeneity can be further explored by calculating 
$F_{s}(k,t)$ at wavenumber corresponding to solute-solvent average
separation. This corresponds to $k^* = 2\pi/\sigma_{12}$, where 
$\sigma_{12} = {1\over 2}(\sigma_{1} + \sigma_{2})$.
In figure 6 we plot the self dynamic structure factor of the solute for 
different mass ratio $M_{R}$, at reduced wavenumber 
$k^* \sim 2\pi/\sigma_{12}$. This is primarily the wavenumber probed by
the solute. At this wavenumber one observes more stretching 
of $F_{s}(k,t)$ at longer times, for all the mass ratios. This is in
agreement with the above argument that since the time window probed by the 
solute is smaller at higher $k$, it probes even larger heterogeneity. 

The emergence of the plateau between the two stretched exponentials in $F_{s}(k,t)$ 
of the solute (figure 6), can be attributed to the separation of time scale of the 
binary collision and the solvent density mode contribution to the particle 
motion. This separation of time scale increases as the solvent mass is 
increased. As the decay after the plateau is mainly due to the density 
mode contribution, the plateau becomes more prominent as the mass of 
the solvent is increased. Table I clearly shows the separation of time
scales where the value of the time constants and the exponents obtained from 
the two different stretched exponential fits to $F_{s}(k,t)$ are presented for 
different mass ratio $M_{R}$.    

The $F_{s}(k,t)$ of each of the individual solute particle obtained from
a single MD run is shown in figure 7, at $k^* \sim 2\pi/\sigma_{12}$ and for 
$M_{R}$ = 250. We find that not only all of them have different time scale 
of relaxation but each of them shows considerable stretching (the 
exponent $\beta \simeq 0.6$) at longer time. This confirms further that each of 
the solute particle probes heterogeneous structure and dynamics of the solvent. 

In table II we present the scaled average (over all the solute particles and 
three independent MD runs) diffusion value of the solute particles obtained 
from the slope of the mean square displacement (MSD) in the diffusive limit, 
for different mass ratio, $M_{R}$. The values of the solute diffusion 
decreases as the mass of the solvent is increased, as expected. The mass
dependence can be fitted to a power law as clearly manifested in figure 8
where we have plotted ln $1/D_{2}$ against ln $m_{1}/m_{2}$, where $D_{2}$ is 
the self-diffusion of the solute particles. The slope of the line is about 0.13.
The small value of the exponent is clearly an indication of the weak mass
dependence of the self-diffusion coefficient of small solute particle on the 
mass of the bigger solvent particles.

Interestingly, it is to be noted that a similar weak power law mass 
dependence was seen in the self-diffusion coefficient of a tagged particle 
on its mass -- the exponent was often found to be around 0.1. Recently, a 
self-consistent mode coupling theory (MCT) analysis successfully explained
this weak mass dependence.$^{23}$    

In figure 9 we plot the normalized velocity autocorrelation 
function $C_v (t)$ of the solute particles for the different
values of $M_{R}$. The velocity correlation function shows
highly interesting features at larger mass ratio. 
Not only does the negative dip becomes 
larger, there develops a second minimum or an extended negative plateau which becomes 
prominent as the mass of the solvent is increased. 
Interestingly, as can be seen from the figure the $C_v(t)$ shows an oscillatory behavior that 
persists for long period. This is clearly evidence for the 'dynamic cage' formation in which 
the solute particle is seen to execute a damped oscillatory motion. Because of the increasing 
effective structural rigidity of the neighboring solvent particles as the mass
of the solvent increases, the motion of the solute particle can be modelled
as a damped oscillator which is reminiscent of the behavior observed in
a deeply supercooled liquid near the glass transition temperature.$^{24}$

The understanding of the microscopic origin of the development of an increasingly 
negative dip followed by pronounced oscillations at longer times in the velocity
autocorrelation function of a supercooled liquid is a subject of much current interest.
The novel molecular-dynamics simulation study of Kivelson and his coworkers$^{25}$ 
surely provide a step forward in this direction. Their simulation study had shown that
the single particle velocity autocorrelation function could be thought of as a sum of
the local rattling motion relative to the center of mass of the neighboring cluster 
and the motion of the center of mass of that cluster. Furthermore, it was observed
that the rich structure displayed by the velocity autocorrelation function at high 
density arise primarily from the relative motion of the tagged particle, that is, 
the rattling motion within the cage formed by the neighboring particles.

\section{Mode Coupling Theory Analysis}

Mode coupling theory remains the only quantitative fully microscopic theory for self-diffusion in
strongly correlated systems. In this section we present a mode coupling theory 
calculation of the solute diffusion for different solvent-to-solute mass ratio $M_{R}$. 
Diffusion coefficient of a tagged solute is given by the well known Einstein relation,

\begin{equation}
D_2=k_{B}T/m_{2}\zeta_2(z=0),
\end{equation}

\noindent
where $D_2$ is the diffusion coefficient of the solute and $\zeta_2(z)$ is the 
frequency dependent friction. $m_2$ is the mass of the solute particle.
Mode coupling theory provides an expression of the frequency dependent  
friction on an isolated solute in a solvent.
 
In the normal liquid regime (in the absence of hopping transport) it can be 
given by,$^{11,12}$ 

\begin{equation}
\frac{1}{\zeta_{2}(z)} \; = \frac{1}{\;\zeta_{2}^{B}(z) \; + \; R_{21}^{\rho \rho }(z)} \:
+R_{21}^{T T }(z) 
\end{equation}   

\noindent
where $\zeta_{2}^{B}(z)$ is the binary part of the friction,
$R_{21}^{\rho \rho }(z)$ is the friction due to the coupling of the
solute motion to the collective density mode of the solvent and  
$R_{21}^{T T }(z)$ is the contribution to the diffusion (inverse of friction) 
from the current modes of the solvent.

 For the present system, we have neglected the contribution from the current 
term, $R_{21}^{T T}(z)$ which is expected to be reasonable at high density and low
temperature. Thus the total frequency 
dependent friction can be approximated as,

\begin{equation}
\zeta_{2}(z) \; \simeq \;\zeta_{2}^{B}(z) \; + \; R_{21}^{\rho \rho }(z) \:
\end{equation} 
 
 The expression for the time dependent binary friction $\zeta_{2}^{B}(t)$, 
for solute-solvent pair, is given by,$^{11,12}$

\begin{equation}
\zeta_{2}^{B}(t)\;=\;\omega^{2}_{o12} exp(-t^{2}/\tau^{2}_{\zeta}),
\end{equation}

\noindent
where $\omega_{o12}$ is now the Einstein frequency of the solute in 
presence of the solvent and is given by,

\begin{equation}
\omega^{2}_{o12}\;=\;\frac{\rho}{3m_{2}}\;\int\;d{\bf r} g_{12}(r){\nabla
}^{2}v_{12}(r).
\end{equation}

\noindent
Here $g_{12}(r)$ is the partial solute-solvent radial distribution function.

In Eq. 6, the relaxation time $\tau_{\zeta}$ is determined from the 
second derivative of $\zeta_{2}^{B}(t)$ at $t=0$ and is given 
by,$^{12,23}$

\begin{eqnarray}
\omega^{2}_{o12}/\tau^{2}_{\zeta}\;&=&\;(\rho/ 6 m_{2} \mu)\int d{\bf r}(\nabla^{\alpha}
\nabla^{\beta}v_{12}({\bf r}))g_{12}({\bf r}) \nonumber\\
&&\times\,(\nabla^{\alpha}\nabla^{\beta}v_{12}({\bf r})) + (1/6\rho)\int[d{\bf k}/(2\pi)^{3}] \nonumber\\
&&\times\,\gamma^{\alpha\beta}_{d12}({\bf k})(S(k)-1)\gamma^{\alpha\beta}_{d12}({\bf k})
\end{eqnarray}

\noindent
where summation over repeated indices is implied. $\mu $ is the reduced mass of 
the solute-solvent pair. Here $S(q)$ is the
static structure factor which is obtained from the HMSA scheme.$^{26}$
The expression for $\gamma^{\alpha\beta}_{d12}({\bf k})$
is written as a combination of the distinct parts of the second moments 
of the longitudinal and transverse current correlation 
functions $\gamma^{l}_{d12}({\bf k})$ and $\gamma^{t}_{d12}({\bf k})$, 
respectively.

\begin{eqnarray}
\gamma^{\alpha\beta}_{d12}({\bf k}&)&\;=\;-(\rho/m_{2} )\int d{\bf r}\;exp(-i{\bf
k}.{\bf r}) g_{12}({\bf r})\nabla^{\alpha}\nabla^{\beta}v_{12}({\bf r}) \nonumber\\
&&\;=\;\hat{k}^{\alpha}\hat{k}^{\beta}\gamma^{l}_{d12}({\bf k})+(\delta_{\alpha\beta}-\hat{k}^{\alpha}\hat{k}^{\beta})\gamma^{t}
_{d12}({\bf k}) 
\end{eqnarray}

\noindent
where  $\gamma^{l}_{d12}({\bf k})=\gamma^{zz}_{d12}({\bf k})$
and $\gamma^{t}_{d12}({\bf k})=\gamma^{xx}_{d12}({\bf k})$. 

The expression for $R_{21}^{\rho\rho}(t)$, for solute-solvent pair, can be 
written as,$^{12,23}$ 

\begin{eqnarray}
R_{21}^{\rho \rho }(t) \;&=&\; \frac{\rho k_{B} T}{m_{2}} \; \int [d{\bf
k}^{\prime}/(2 \pi)^{3}] (\hat{k}.{\hat{k}}^{\prime})^{2} {k^{\prime}}^{2}
[c_{12}(k^{\prime})]^{2} \nonumber\\ 
&&\times\,[F^{s}(k^{\prime},t)F(k^{\prime},t) - F^{s}_{o}(k^{\prime},t) 
F_{o}(k^{\prime},t)]
\end{eqnarray}

 In Eq. 10, $c_{12}(k)$ is the two particle (solute-solvent) direct correlation
in the wavenumber $(k)$ space which is obtained here from the HMSA scheme.$^{26}$ 
The partial radial distribution function ($g_{12} (r)$) required to 
calculate the Einstein frequency ($\omega_{o12}$) and the binary time
constant ($\tau_{\zeta}$) is obtained from the present simulation study.
$F(k,t)$ is the intermediate scattering function of the solvent, and 
$F_{o}(k,t)$ is the inertial part of the intermediate scattering function.
$F_{s}(k,t)$ is the self-intermediate scattering function of the solute
and $F_{o}^s(k,t)$ is the inertial part of $F_{s}(k,t)$. 

It should be noted here that the short time dynamics of the density term
used in Eq.10, is different from the conventional mode coupling formalism.$^{27}$
This prescription has recently been proposed to explain the weak power law 
mass dependence of the self-diffusion coefficient of a tagged particle.$^{23}$  
The detailed discussion on this prescription has been given elsewhere.$^{12,23}$  
 
 Since the solvent is much heavier than the solute, the decay of solvent dynamical 
variables are naturally much slower than those of the solute. Since the decay of 
$F_{s}(k,t)$ is much faster than $F(k,t)$, in Eq. 10 the contribution
from the product, $F_{s}(k,t)F(k,t)$ mainly governed by the time scale of 
decay of $F_{s}(k,t)$. Thus the long time part of the $F(k,t)$ becomes unimportant 
and the viscoelastic expression$^{12}$ for $F(k,z)$ (Laplace transform of 
$F(k,t)$) obtained by using the well-known Mori continued-fraction expansion, truncating at 
second order would be a reasonably good approximation. The expression of $F(k,z)$, can 
be written as,$^{11,12}$
\begin{equation}
F(k,z)\;=\;\frac{S(k)}{z \; + \; \frac {<{\omega_{k}}^{2}>}{z\; +
\;\frac{\Delta_{k}}{z\; + \;{\tau_{k}}^{-1}}}},
\end{equation}
\noindent
where $F(k,t)$ is obtained by Laplace inversion of $F(k,z)$, the dynamic 
structure factor. Because of the viscoelastic approximation, $<\omega_{k}^2>$ and 
$\Delta_{k}$ and also $\tau_{k}$ are determined by the static pair correlation 
functions. The static pair correlation functions needed are the static structure
factor $S(q)$ and the partial solvent-solvent radial distribution function $g_{11}(r)$.
$S(q)$ is obtained by using the HMSA scheme$^{26}$ and $g_{11}(r)$ is taken
from the present simulation study.
 
 We have used the recently proposed$^{12,23}$ generalized self-consistent scheme
to calculate the friction, $\zeta(z)$, which makes use of the well-known Gaussian 
approximation for $F_{s}(k,t)$,$^{28}$
\begin{eqnarray}
F^{s}(k,t&)&\;=\;\exp (-\frac{k^{2} <\Delta r^{2}(t)>}{6}) \nonumber\\
&&\;=\;\exp\Biggl[ - {k_BT\over m_2} k^2 \int ^{t}_{0} d\tau C_{v}(\tau) (t-\tau)\Biggr]
\end{eqnarray}
\noindent
where $<\Delta r^{2}(t)>$ is the mean square displacement (MSD) and $C_v(t)$ is
the time-dependent velocity autocorrelation function (VACF) of the solute particles. 
The time-dependent VACF is obtained by numerically Laplace inverting the frequency-dependent
VACF, which is in turn related to the frequency-dependent friction through the following
generalized Einstein relation given by
\begin{equation}
C_{v}(z)= \frac {k_{B}T} {m_s(z +\zeta(z))}
\end{equation}
\noindent
Thus in this scheme the frequency-dependent friction has been calculated self-consistently
with the MSD. The details of implementing this self-consistent scheme is given
elsewhere.$^{12,23}$

 We have evaluated the diffusion coefficient $D_2$ by using the above mentioned 
self-consistent scheme. The calculated diffusion value was found to be higher than the 
simulated one. This may be partly due to the observed faster decay of calculated 
$F_{s}(k,t)$ than the simulated one. This in turn could be due to the Gaussian approximation 
for $F_{s}(k,t)$ which truncates the cumulant expression of $F_{s}(k,t)$ beyond the
quadratic ($k^2$) term.$^{29}$ 
However, the higher order terms which are the systematic corrections to the Gaussian 
forms can be increasingly important at intermediate times and wave numbers ($k$).$^{28}$ 
               
 Therefore, we have performed MCT calculations using the {\it simulated} $F_{s}(k,t)$ 
evaluated at different values of the wave number, $k = nk_{min}$, where 
$k_{min} = 2\pi/{\bar L}$ (${\bar L}$ stands for the average size of the simulation 
cell) and $n$ is an integer varied in the range, $1\le n \le 35$. $F_{s}(k,t)$ is then 
obtained by interpolating the simulated $F_{s}(k,t)$ so obtained at different wave numbers.
The calculated value of the binary term and the density term contribution
to the friction for different values of $M_{R}$ are presented in table III.

Earlier mode coupling theoretical calculations$^{11}$ for small solute 
was performed by keeping the solute and the solvent mass {\it equal}. In those calculations 
it was found that for size ratio $S_{R}$ = 5, the solute motion is primarily 
determined by the binary collision between the solute and the solvent particles.
It was also shown that due to the decoupling of the solute motion from the structural 
relaxation of the solvent, the contribution of the density mode of the solvent was much 
smaller than that of the binary term. In the present calculation we find that due to 
smaller mass of the solute, the contribution of the binary term decreases
with increase in the solvent mass as is clearly evident from table III.
 
When compared to the simulated diffusion values, it is clearly seen from table II that 
although the MCT qualitatively predicts the diffusion value for $M_{R}$ = 5, it breaks 
down at large values of the mass ratios. This may be because MCT overestimates the 
friction contribution from the density mode. This breakdown of the MCT for large mass 
ratio can be connected to its breakdown observed near the glass transition temperature. 
For large solvent mass, the system is almost frozen and the dynamic structure factor 
of the solvent decays in a much longer time scale when compared to the solute. 
So from the point of view of the solute, it probes an almost quenched system which can be 
expected to show the behavior very similar to a system, near the glass transition. Just
as near the glass transition temperature, the hopping mode also plays the dominant
role in the diffusion process.

\section{Effect of Dynamic Heterogeneity on $F_{s}(k,t)$ of the Solute}

It is well-known that the self-intermediate scattering function, $F_{s}(k,t)$ can be 
formally expressed by the following cumulant expansion in powers of $k^2$,$^{29}$
\begin{eqnarray}
F_s(k,t)\;&=&\;\exp(-{1\over 6}k^2<\Delta r^2(t)>)\Biggl[ 1 + {1\over 2}\alpha_2(t) \nonumber\\
&&\times\,({1\over 6}k^2<\Delta r^2(t)>)^2 + O(k^6)\Biggr]
\end{eqnarray}
\noindent
where $\alpha_2(t)$ is defined as
\begin{equation}
\alpha_2(t) = {3 <\Delta r^4(t)>\over 5 <\Delta r^2(t)>^2} - 1
\end{equation}
\noindent
In the stable fluid range, it has been generally found that the Gaussian
approximation to $F_{s}(k,t)$, the leading term in the above cumulant 
expansion, provide a reasonably good description of the dynamics of the 
system. The higher order terms which are the systematic corrections to the 
Gaussian approximation in the cumulant expansion are found to be small.

However, in a supercooled liquid, this is not the case. The dynamical
heterogeneities observed in a deeply supercooled liquid are often 
manifested as the magnitude of the deviation of $\alpha_2(t)$, the so-called 
non-Gaussian parameter, from zero. It has been observed that $\alpha_2(t)$ 
deviates more and more strongly and decays more and more slowly with increase 
in the degree of supercooling.$^{30}$ In the present study, a similar 
behavior has also been observed in $\alpha_2(t)$ (calculated for the solute). 
The height of the maximum in $\alpha_2(t)$ increases as the mass of the solvent is 
increased (see figure 5), which is clear evidence that the solute probes increasingly
heterogeneous dynamics.

It is generally believed that the dominant corrections to the Gaussian result
are provided by the term containing $\alpha_2(t)$. Thus, it would be interesting
to see whether this term alone is sufficient to explain the observed stretching 
in $F_{s}(k,t)$ of the solute at longer time. In order to quantify this, we have plotted in 
figure 10 the simulated $F_{s}(k,t)$ along with the $F_{s}(k,t)$ obtained after 
incorporating the lowest order correction ($k^4$ term) to the Gaussian approximations 
for mass ratio, $M_{R}$ = 50 at the reduced wavenumber $k^* \sim 2\pi/\sigma_{12}$. 
For comparison, the Gaussian approximation to $F_{s}(k,t)$ is also shown. It is clearly 
seen that the first non-Gaussian correction to $F_{s}(k,t)$ is not sufficient to 
describe the long time stretching predicted by the simulation. This clearly 
indicate that at the length scales probed by the solute, {\it the higher order 
corrections cannot be neglected}. It is the nearly quenched inhomogeneity probed by 
the solute particles over small length scales which play an important role in the 
dynamics of the system. The stretching of $F_{s}(k,t)$ observed in simulation could
be intimately connected with this nearly quenched inhomogeneity probed by the solute
particles. 

\section{Conclusions}

 Let us first summarize the main results of this study.
We have investigated by using the molecular dynamics simulation
the diffusion of small light particles in a solvent composed of larger massive 
particles for a fixed solvent-to-solute size ratio ($S_{R}$ = 5) but with a large 
variation in mass ratio (where the mass of the solute is kept constant). 
In addition, a mode-coupling theory (MCT) analysis of diffusion is also presented.
It is found that the solute dynamics remain surprisingly coupled to the solvent 
dynamics even in the limit of highly massive solvent. Most interestingly, with 
increase in mass ratio, the self-intermediate scattering function of the solute develops 
a stretching at long time which, for intermediate values of mass ratio, 
could be fitted to a single stretched exponential 
function with the stretching exponent, $\beta \simeq 0.6$. 
In the limit of very large mass ratio, 
the existence of two stretched exponential separated by
a power law type plateau is observed. This behavior is found to arise from 
increasingly heterogeneous environment probed by the solute particle as one 
increases the mass of the solvent particles. The MCT calculation of self-diffusion 
is found to agree qualitatively with the simulation results for small mass ratio. 
However, it fails to describe the simulated prediction at large mass ratios. 
The velocity correlation function of the solute shown interesting oscillatory
structure.

Several of the results observed here are reminiscent of the relaxation of the self-intermediate 
scattering function, $F_{s}(k,t)$ observed in the deeply supercooled
liquid near its glass transition temperature. In that case also, one often observes
combination of power-law and stretched exponential in the decay of the intermediate
scattering function. We find that 
even the breakdown of MCT at large mass ratio could be connected 
to its breakdown near the glass transition temperature because it is the neglect of the
spatial hopping mode of particles which is responsible for the breakdown of MCT.
It is to be noted that those hoppings which are mostly ballistic 
in nature (after a binary collision) have already been incorporated in MCT. However, MCT 
does not include the hoppings which involve collective displacement involving several 
molecules.$^{15}$ 

It should be pointed out that in the MCT calculation, we have neglected the
contribution of the current term. While the current contribution may improve 
the agreement between the simulation and MCT result for small mass ratio ($M_{R}$ = 5),
its contribution at larger mass ratio is not expected to change the results
significantly, because the discrepancy is very large.    

 The origin of the power law remains to be investigated in more detail. Our 
preliminary analysis shows that this may be due to the separation of the time scale
between the first weakly stretched exponential (due to the dispersion in the binary-type
interaction term) and the second, later more strongly stretched exponential (which is due
to the coupling of the solute's motion to the density mode of the slow solvent).
This separation arises because these two motions are very different in nature.
However, a quantitative theory of this stretching and power-law is not available 
at present.

While the origin of the stretching of $F_{s}(k,t)$ can be at least qualitatively understood
in terms of the inhomogeneity experienced by the solute, the origin of hopping is less clear.
In the supercooled liquid, hopping is found to be correlated with anisotropic local stress$^{15}$
which is unlikely in the present system which is at lower density and pressure. 

 Finally we note that the system investigated here is a good candidate to understand
qualitative features of relaxation in a large variety of systems, such as concentrated
solution of polysaccharide in water and also motion of water in clay.

{\bf Acknowledgments}

This work was supported in part by the Council of Scientific and Industrial
Research (CSIR), India and the Department of Science and Technology (DST), India.
One of the authors (R.K.M) thanks the University Grants Commission (UGC) for
providing the Research Scholarship.

\begin{center}
{\bf TABLE I.}
\end{center}
\noindent The time constants ($\tau_1$ and $\tau_2$) and the exponents ($\beta_1$ and $\beta_2$)
obtained from the stretched exponential fit to the $F_{s}(k,t)$ at the reduced wavenumber
$k^* \sim 2\pi/\sigma_{12}$ for different solvent-to-solute
mass ratio ($M_{R}$). 
\vspace{0.5cm}
\begin{center}
\begin{tabular}{|c|c|c|c|c|} \hline
${\bf M_R = {m_1\over m_2}}$& ${\tau_1}$& ${\beta_1}$& ${\tau_2}$& ${\beta_2}$ \\ \hline
&&&&\\
5 &0.082 &0.70 & & \\
&&&&\\ \hline
&&&&\\
25 &0.08 &0.96 &0.49 &0.67 \\
&&&&\\ \hline 
&&&&\\
50 &0.083 &0.94 &0.59 &0.64 \\
&&&&\\ \hline
&&&&\\
250 &0.085 &0.91 &1.01 &0.635 \\
&&&&\\ \hline
\end{tabular}
\end{center}

\begin{center}
{\bf TABLE II.}
\end{center}
\noindent The self-diffusion coefficient values of the solute particle
predicted by the simulation and obtained from the MCT calculations
for different solvent-to-solute mass ratio ($M_{R}$).  
\vspace{0.5cm}
\begin{center}
\begin{tabular}{|c|c|c|} \hline
${\bf M_R = {m_1\over m_2}}$& ${\bf D_2^{sim}}$& ${\bf D_2^{MCT}}$ \\ \hline
&&\\
5 &0.135 &0.1065 \\
&&\\ \hline
&&\\
25 &0.108 &0.0675 \\
&&\\ \hline
&&\\
50 &0.101 &0.053  \\
&&\\ \hline
&&\\
250 &0.0805 &0.032 \\
&&\\ \hline
\end{tabular}
\end{center}

\begin{center}
{\bf TABLE III.}
\end{center}
\noindent The contribution of the binary (${\zeta_{2}^{B}}$) and the density 
mode (${R_{21}^{\rho\rho}}$) of the friction for different 
solvent-to-solute mass ratio ($M_{R}$). 
\vspace{0.5cm}
\begin{center}
\begin{tabular}{|c|c|c|} \hline
${\bf M_R = {m_1\over m_2}}$& ${\bf \zeta_{2}^{B}}$& ${\bf R_{21}^{\rho\rho}}$ \\ \hline
&&\\
5 &23.65 &13.95 \\
&&\\ \hline
&&\\
25 &25.4 &33.8 \\
&&\\ \hline
&&\\
50 &25.45 &50.2 \\
&&\\ \hline
&&\\
250 &25.58 &99.8 \\
&&\\ \hline
\end{tabular}
\end{center}
   
\newpage 
{\large \bf Figure Captions}
\vspace{0.5cm}

{\bf Figure 1.} The time dependence of the displacements for a solute particle
at different solvent-to-solute mass ratio, $M_{R}$: (a) $M_{R}$ = 5, (b) $M_{R}$ = 25, 
(c) $M_{R}$ = 50, and (d) $M_{R}$ = 250. Note that the time is scaled by 
$\sqrt{m\sigma_{11}^2/k_BT}$. The time unit is equal to 2.2 ps if argon units
are assumed.

{\bf Figure 2.} The same plot as in figure 1, but for a solvent particle at different
$M_{R}$: (a) $M_{R}$ = 5, (b) $M_{R}$ = 25, (c) $M_{R}$ = 50, and (d) $M_{R}$ = 250.
Note that here also the time is scaled by $\sqrt{m\sigma_{11}^2/k_BT}$.   

{\bf Figure 3.} The self-intermediate scattering function $F_{s}(k,t)$ for the solute
particles for different mass ratio $M_{R}$, at reduced wavenumber 
$k^* = k\sigma_{11} \sim 2\pi$. The solid line represents for $M_{R}$ = 5, the dashed 
line for $M_{R}$ = 25, the dotted line for $M_{R}$ = 50, and the dot-dashed line 
for $M_{R}$ = 250. Note the stretching in $F_{s}(k,t)$ at longer time with increase 
in $M_{R}$. For details, see the text.

{\bf Figure 4.} The self-intermediate scattering function $F_{s}(k,t)$ for the solvent
particles for different mass ratio $M_{R}$, at reduced wavenumber $k^* \sim 2\pi$.
The solid line represents for $M_{R}$ = 5, the dashed line for $M_{R}$ = 25, the dotted
line for $M_{R}$ = 50, and the dot-dashed line for $M_{R}$ = 250. 

{\bf Figure 5.} The behavior of non-Gaussian parameter $\alpha_2(t)$ calculated for 
the solute particles at different mass ratio $M_{R}$. The solid line, the dashed line,
the dotted line, and the dot-dashed line are for $M_{R}$ = 5, 25, 50 and 250, respectively.   
   
{\bf Figure 6.} The self-intermediate scattering function $F_{s}(k,t)$ for the
solute particles for different mass ratio $M_{R}$ as in Fig. 3, but at the reduced
wavenumber $k^* \sim 2\pi/\sigma_{12}$. This is primarily the wavenumber probed 
by the solute. Note the emergence of a plateau at larger mass ratio. For details, see
the text.

{\bf Figure 7.} The self-intermediate scattering function $F_{s}(k,t)$ for each of the
ten individual solute particle obtained from a single MD run for mass ratio $M_{R}$ = 250.
They are calculated at the reduced wavenumber $k^* \sim 2\pi/\sigma_{12}$. 

{\bf Figure 8.} The plot of $ln_e 1/D_2$ vs $ln_e m_1/m_2$, where $D_2$ is the self-diffusion
of the solute. $m_1$ and $m_2$ are the masses of the solvent and solute particles, respectively.
The slope of the straight line is about 0.13. This suggests a weak power-law mass dependence
of the solute diffusion on the mass of the bigger solvent particles.    

{\bf Figure 9.} The velocity autocorrelation function $C_v(t)$ for the solute particles
at different values of mass ratio $M_{R}$. The solid line represents for $M_{R}$ = 25, the dotted
line for $M_{R}$ = 50, and the dashed line for $M_{R}$ = 250. The plot shows an increase in the 
negative dip with increase in mass ratio. For details, see the
text. 

{\bf Figure 10.} Comparison of the simulated self-intermediate scattering 
function $F_{s}(k,t)$ of the solute particles with the $F_{s}(k,t)$ obtained 
after incorporating the lowest order correction ($k^4$ term ) to the Gaussian
approximation in the cumulant expansion (Eq. 14). The Gaussian approximation
to $F_{s}(k,t)$ is also shown. The mean-squared displacement ($<\Delta r(t)^2>$) and the
non-Gaussian parameter ($\alpha_2(t)$) required as an input are obtained from the simulation.
The plot is at the reduced wavenumber $k^* \sim 2\pi/\sigma_{12}$ and for the mass ratio
$M_{R}$ = 50. $F_{s}(k,t)$ obtained from the simulation is represented by the solid line, the 
dashed line represents the $F_{s}(k,t)$ obtained after the lowest order correction to the
Gaussian approximation and the dotted line represents the Gaussian approximation. 
For the detailed discussion, see the text.    
\end{multicols}
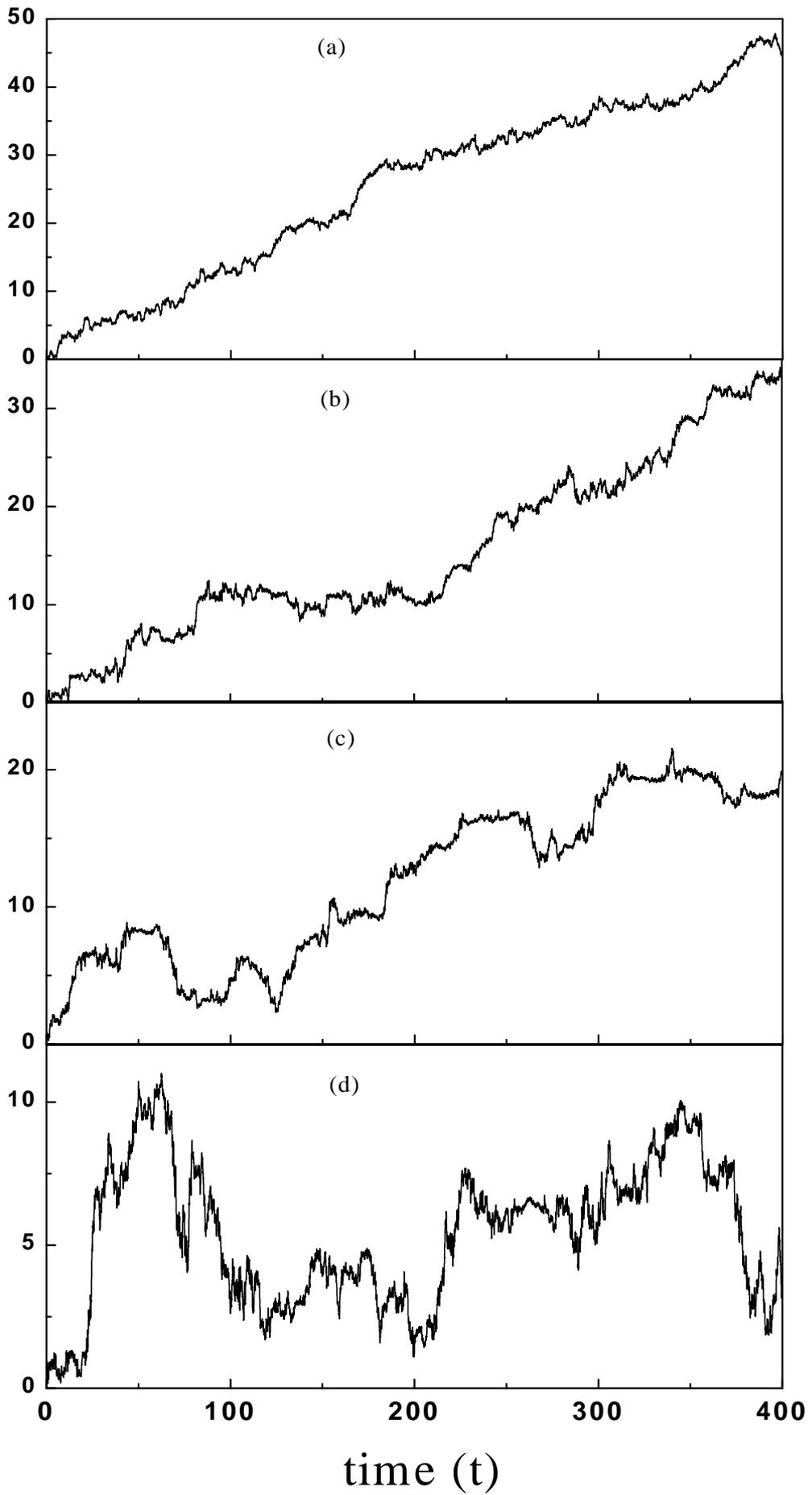
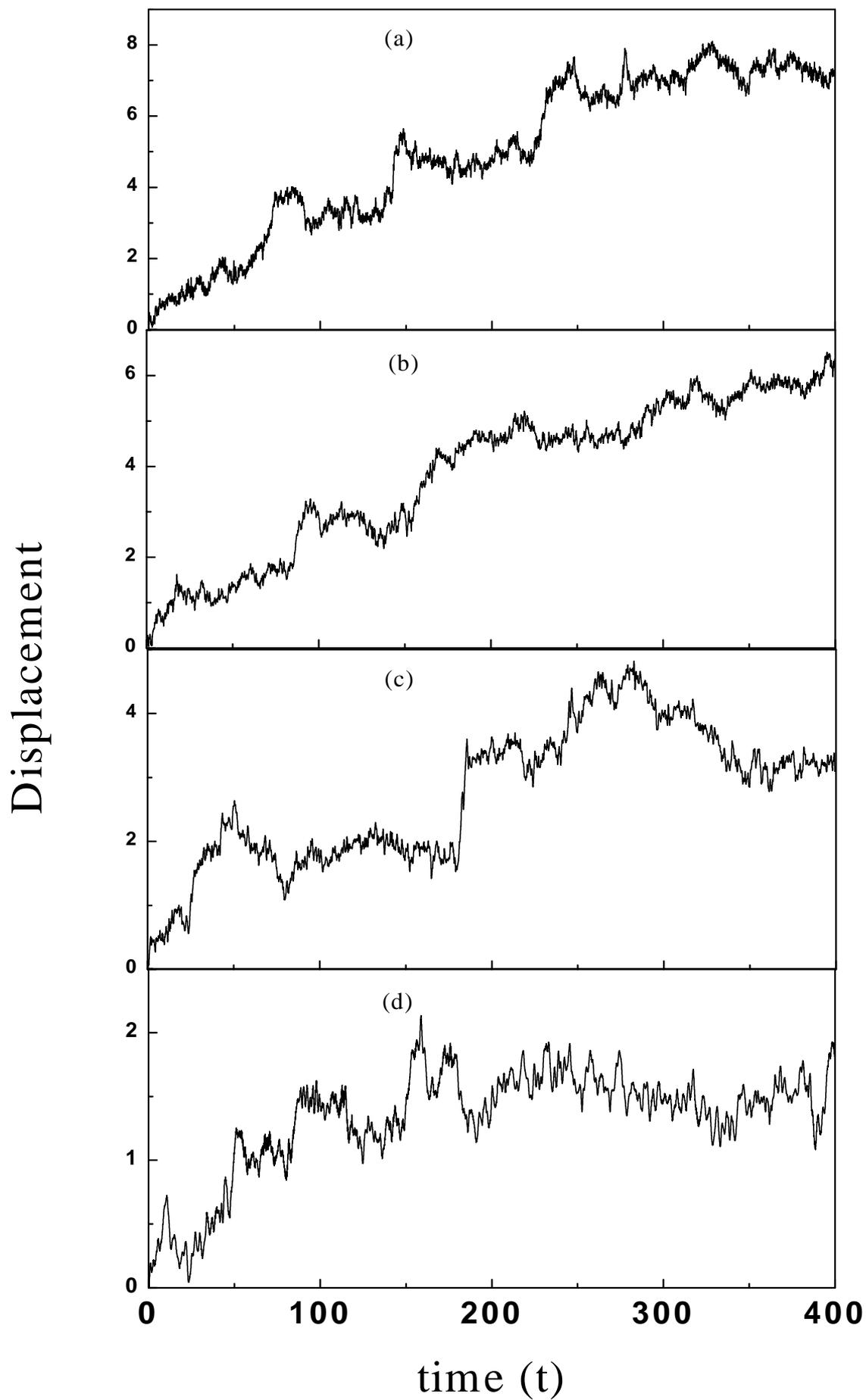
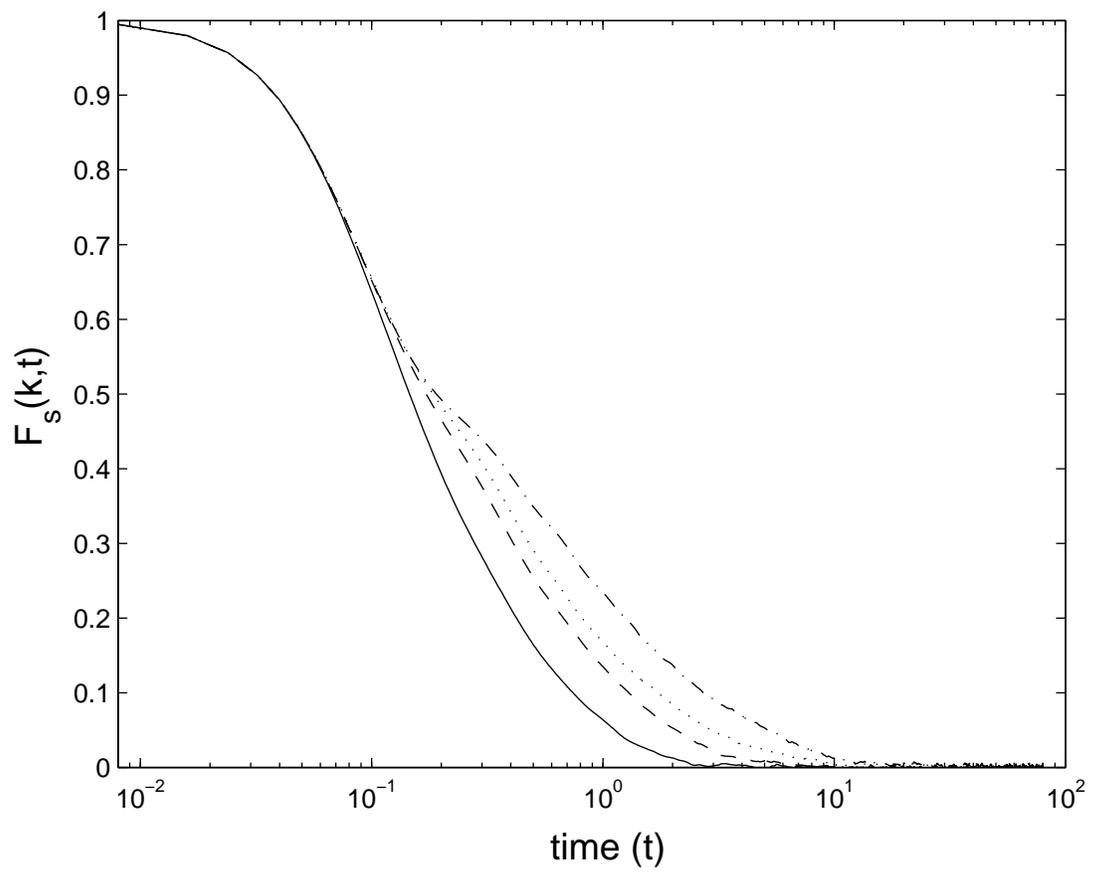
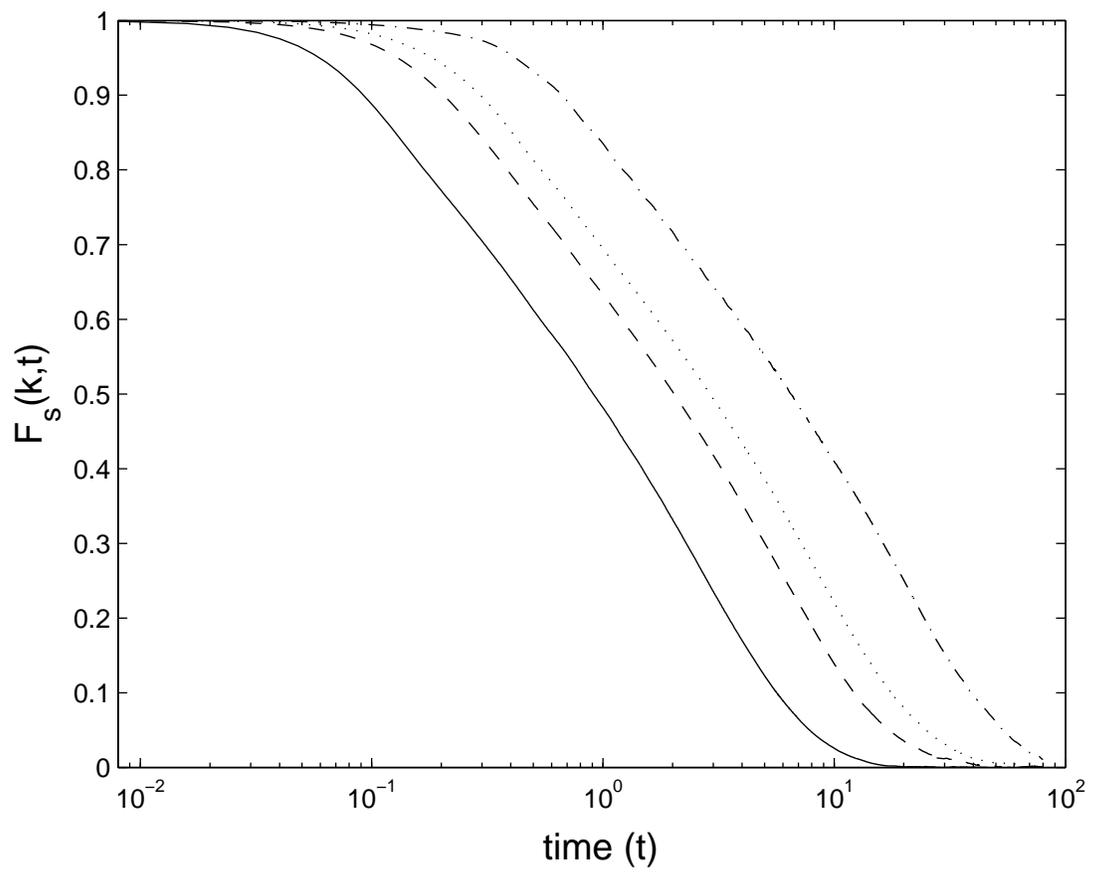
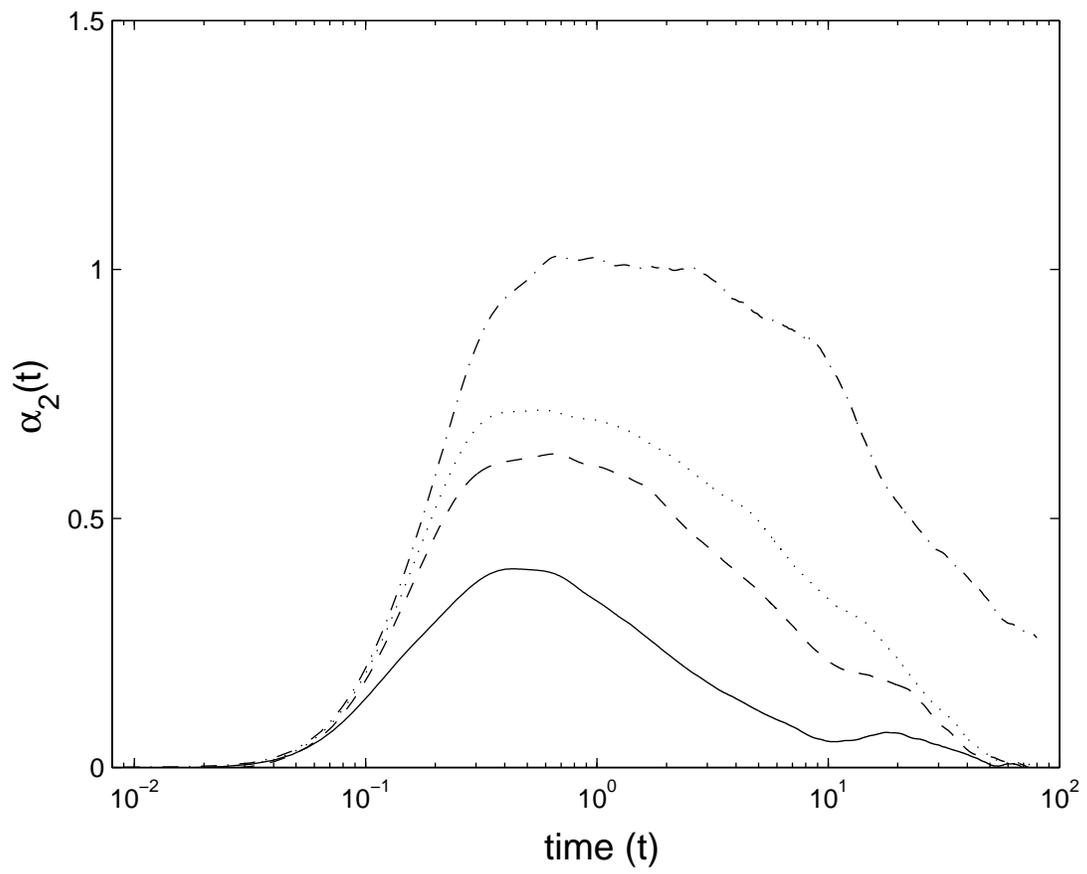
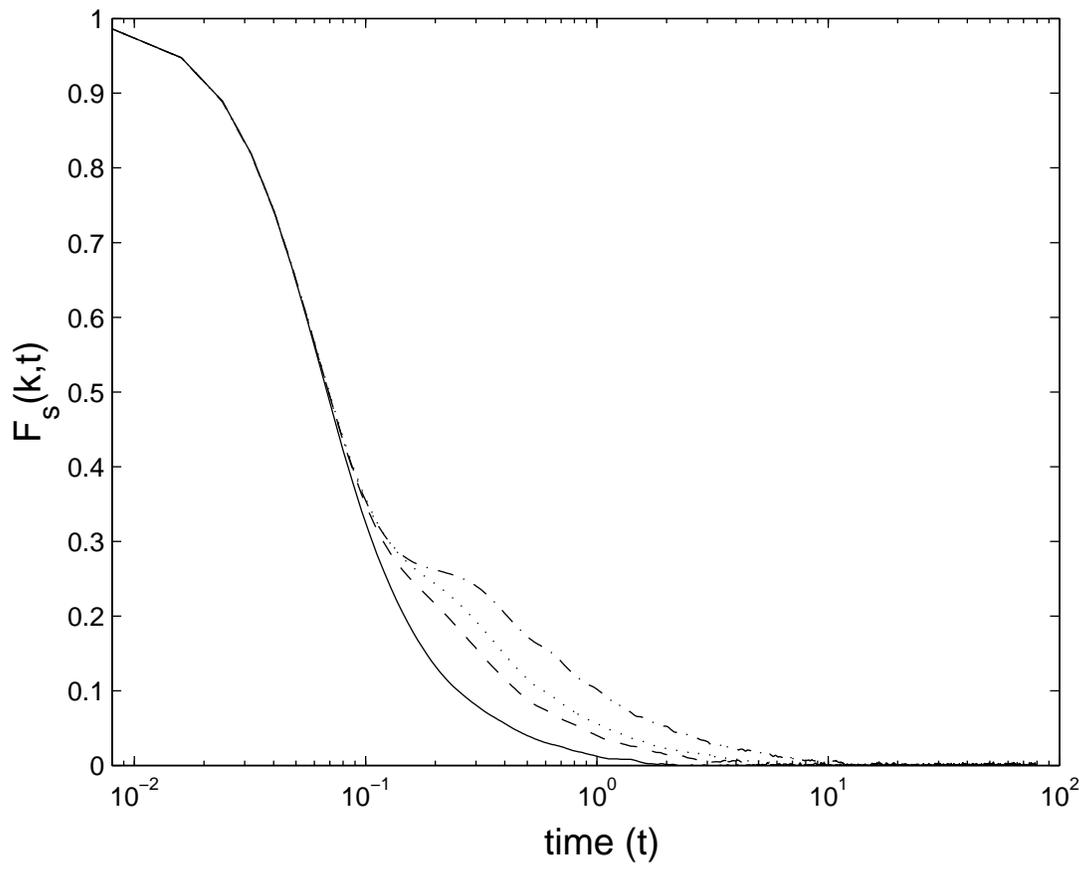
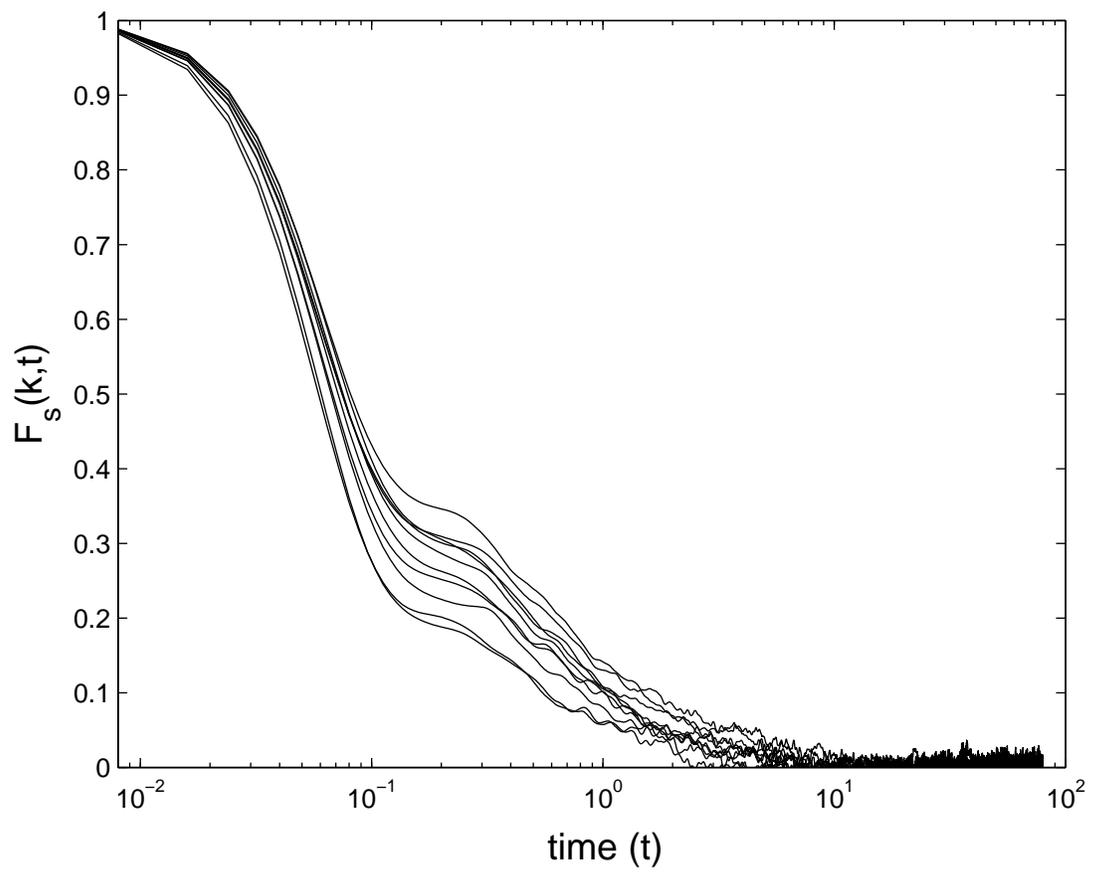
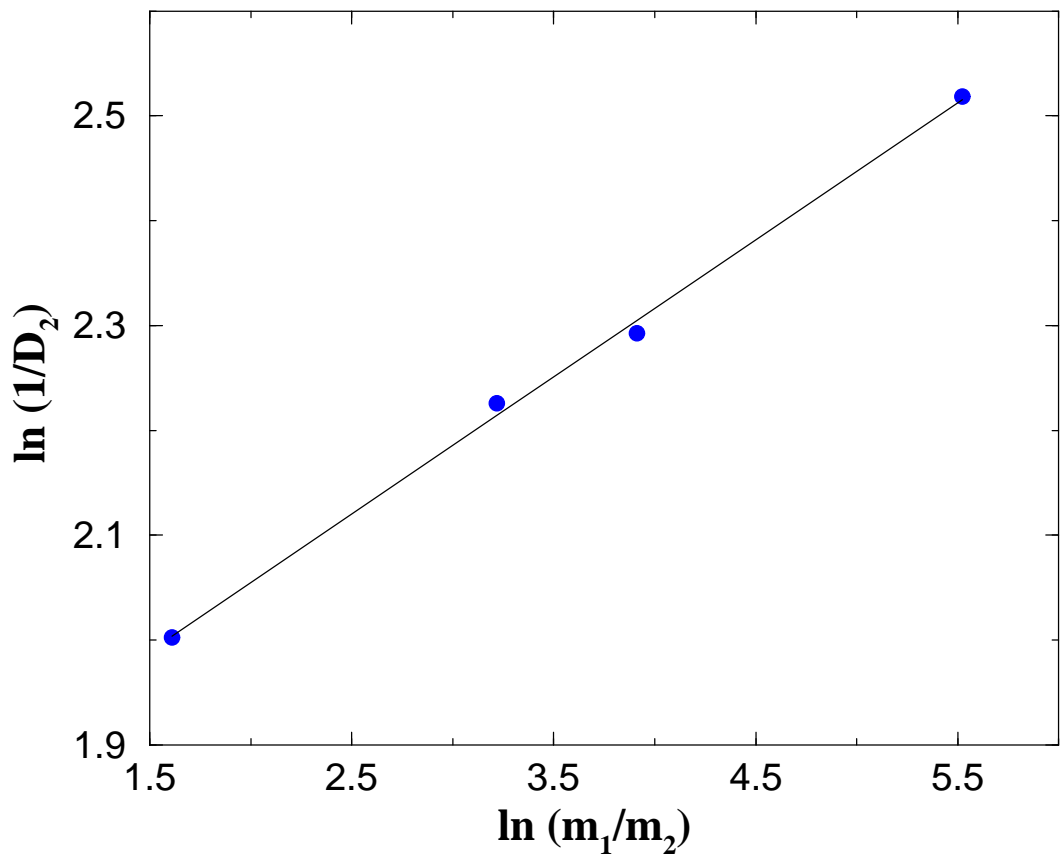
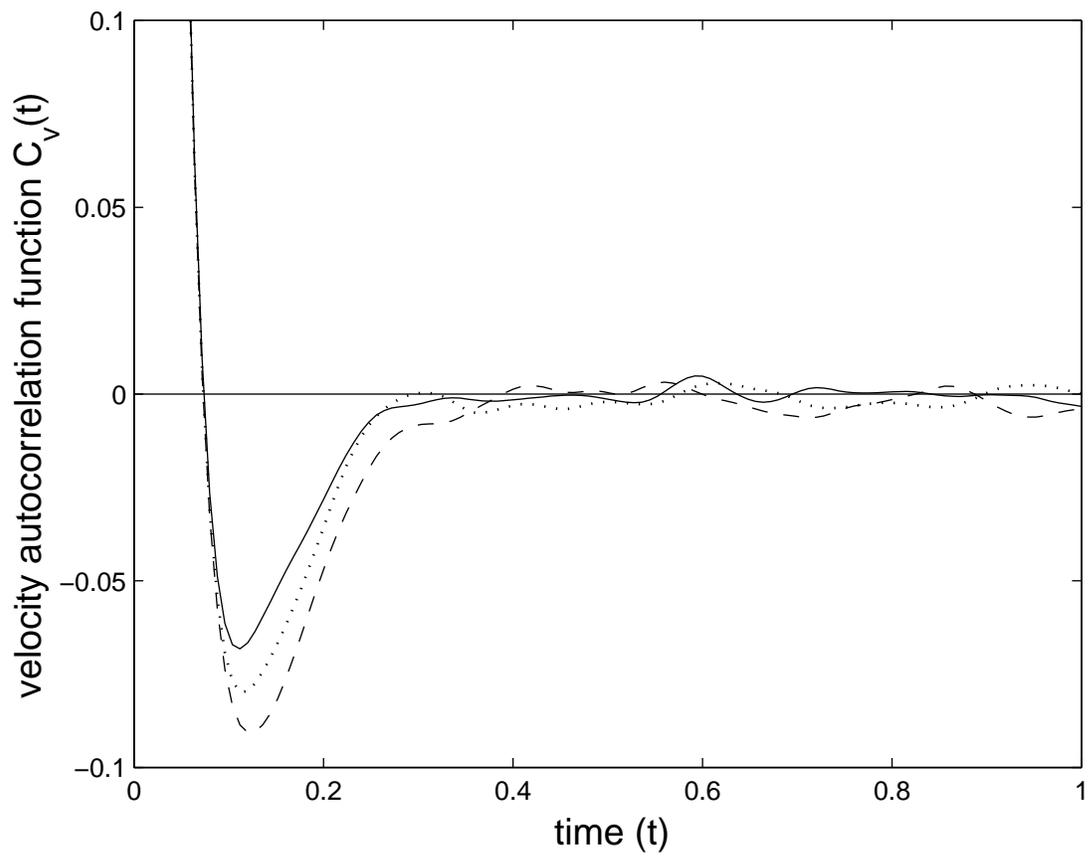
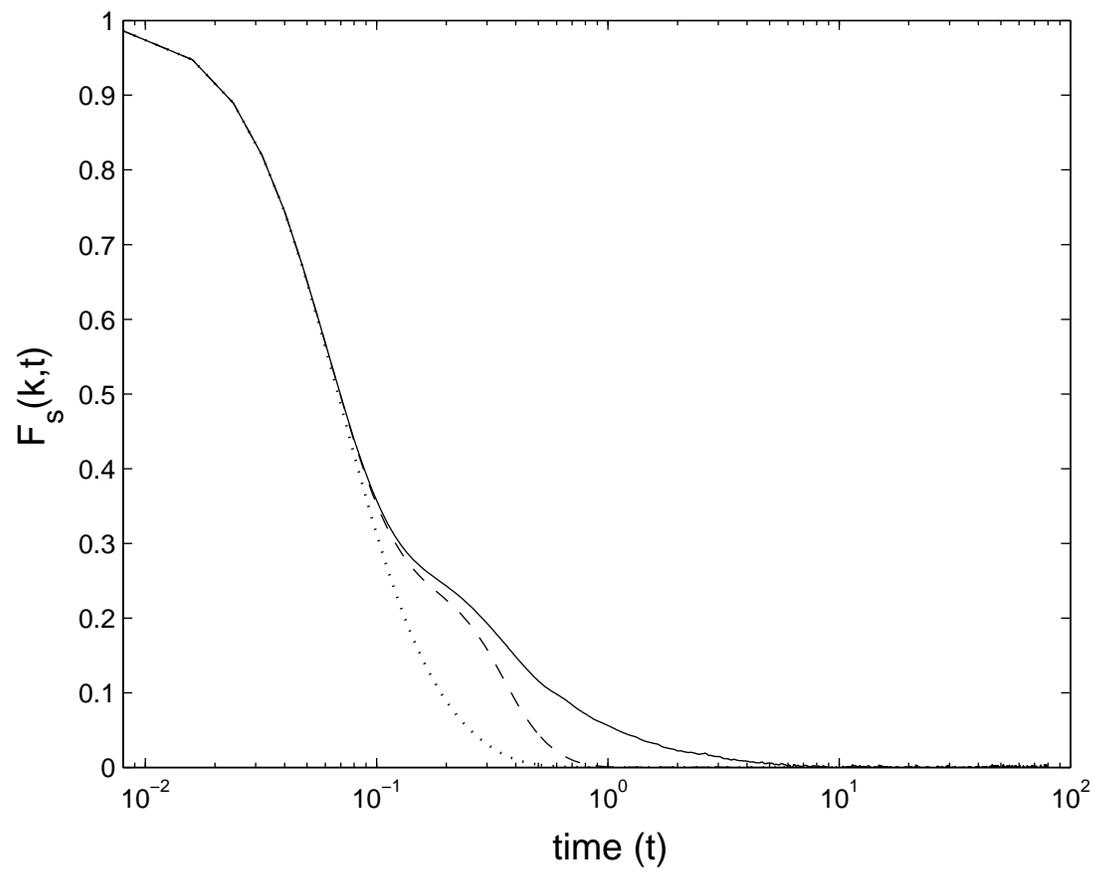
\end{document}